\documentclass[
aps,%
12pt,%
final,%
notitlepage,%
oneside,%
onecolumn,%
nobibnotes,%
nofootinbib,
superscriptaddress,%
noshowpacs,%
centertags]%
{revtex4}
\usepackage[russian]{babel}
\usepackage{epsfig}

\begin{document}





\title{Asymmetry Function of Interstellar Scintillations of Pulsars}

\affiliation{Pushchino Radio Astronomy Observatory, Astrospace
Center of the Lebedev Institute of Physics, Russian Academy of
Sciences, Pushchino, Moscow oblast', 142290 Russia}

\author{\firstname{V.I.}~\surname{Shishov}}

\author{\firstname{T.V.}~\surname{Smirnova}}

\received{February 27, 2004}
\revised{May 18, 2005}

\begin{abstract}
A new method for separating intensity variations of a source's
radio emission having various physical natures is
proposed. The method is based on a joint analysis of the
structure function of the intensity variations and the asymmetry function,
which is a generalization of the asymmetry coefficient and
characterizes the asymmetry of the distribution function of the
intensity fluctuations on various scales for the inhomogeneities in
the diffractive scintillation pattern. Relationships for the
asymmetry function in the cases of a logarithmic normal
distribution of the intensity fluctuations and a normal distribution of
the field fluctuations are derived. Theoretical relationships and
observational data on interstellar scintillations of pulsars
(refractive, diffractive, and weak scintillations) are compared.
Pulsar scintillations match the behavior expected for a normal
distribution of the field fluctuations (diffractive scintillation) or
logarithmic normal distribution of the intensity fluctuations
(refractive and weak scintillation). Analysis of the
asymmetry function is a good test for distinguishing scintillations
against the background of variations that have different origins.
\end{abstract}

\maketitle

\section{INTRODUCTION}

In analyses of interstellar scintillations of intrinsically
variable radio sources, the problem of separating variations of
the source emission with different
origins arises. In particular, this is true of
separating intrinsic variations of pulsar radio emission,
interstellar scintillations, and additive and modulational
noise. A similar problem appears in the analysis of
rapid variability of extragalactic sources. In [1--5], it was
proposed to use the asymmetry coefficient of the distribution
function of the radio-emission intensity fluctuations as a test
for determining the nature of these variations. The
asymmetry coefficient is a measure of the deviation of the
distribution function of the intensity fluctuations from a normal
distribution, and is defined as
\begin{gather}
\gamma  = \frac{\langle (I - \langle I \rangle )^3
\rangle}{[\langle ( I - \langle I \rangle)^2 \rangle]^{3/2}} =
\frac{M_3}{M_2^{3/2}} . \label{eq1:Shishov_n}
\end{gather}
Here, $\langle I \rangle$ is the mean intensity of the source's
radioe emission, and $M_3$ and $M_2$ are the third and second central
moments of the distribution of the intensity fluctuations. In the
case of scintillations, $\gamma$ is positive, and is
related to the scintillation index by the linear relationship
\begin{gather}
\gamma  =  A m , \label{eq2:Shishov_n}
\end{gather}
where $A$ is a numerical factor and $m$ is the scintillation
index, which is defined as
\begin{gather}
m^2  = \frac{\langle (I - \langle I \rangle)^2 \rangle}{\langle I
\rangle^2}  = \frac{M_2}{\langle I \rangle ^2} . \label{eq3:Shishov_n}
\end{gather}

If we are dealing with an extragalactic source consisting of a
compact scintillating component (core) and an extended
non-scintillating component (halo), $I$ corresponds to the flux
from the scintillating component. The factor $A$
depends on the form of the turbulence spectrum and the
scintillation regime. In particular, in the case of weak
scintillations of a pointlike source in the Fraunhofer zone relative
to the outer turbulence scale (i.e., the characteristic size
of the largest inhomogeneities), the flux fluctuations are distributed
according to a Rice--Nakagami law [6, 7], and $A = 3/2$. In the case of
weak scintillations of a pointlike source in the Fresnel zone
relative to the inner scale (i.e., the characteristic size of the
smallest inhomogeneities), the flux fluctuations are distributed
according to a lognormal law [8], and $A = 3$. As is shown in [1],
the relationship (2) is a good test for isolating interstellar
scintillations that give rise to flux variations of extragalactic
radio sources.

However, the asymmetry coefficient is an integrated parameter, which
describes the summed fluctuations on all scales. Interstellar
scintillations can dominate on some time scales, while intrinsic
variations or noise can dominate on others. To analyze the nature of
the fluctuations on various scales, it is desirable to introduce
some function that would yield the value of the asymmetry coefficient
for a given time scale in place of the asymmetry coefficient. Below, we
will introduce such an asymmetry function,
$\gamma_2 (\tau)$, which is a generalization of the asymmetry
coefficient, in the same way that the structure function is a
generalization of the variance of the temporal process. We will also
investigate the application of the asymmetry function to the
analysis of interstellar scintillations of pulsars.

\section{DEFINITION OF THE ASYMMETRY FUNCTION}

Let us introdcue the first and second differences
\begin{gather}
\Delta_1 (\tau)  =  I (t + \tau) -  I(t) ,\label{eq4:Shishov_n} \\
\Delta_2 (\tau)  =  I (t + \tau) - 2 I(t) + I (t - \tau) ,
\nonumber
\end{gather}
where $t$ is the time and and $\tau$ is the time shift from
instant $t$. The structure function of the flux fluctuations is defined as
\begin{gather}
D_I (\tau)  =  \langle [\Delta_1 (\tau) ]^2 \rangle = 2  [\langle
I^2 \rangle - \langle I (t + \tau) I (t) \rangle] \label{eq5:Shishov_n}
\end{gather}
and the mean square of the second difference is
\begin{gather}\label{eq6:Shishov_n}
\langle [ \Delta_2 (\tau) ]^2 \rangle  =  \langle \{ [ I (t +
\tau) - I (t) ] \\
\nonumber{}- [ I (t) - I (t - \tau ) ] \}^2 \rangle =\langle \{ [
I (t + \tau) - I (t) ]^2 \\
\nonumber{}+  [ I (t) - I (t - \tau) ]^2 - 2 [ I (t + \tau) - I(t) ] \\
\nonumber{}\times[ I (t) - I (t - \tau) ] \} \rangle
 =    4 D_I (\tau) -  D_I (2\tau)  .
\end{gather}
We also introduce the third moment of the second difference
$\Delta_2 (\tau)$ using the formula
\begin{gather} \label{eq7:Shishov_n}
\langle [\Delta_2 (\tau)]^3 \rangle  = {-} 6 \langle I^3 \rangle +
6 \langle I^2 (t + \tau) I (t - \tau) \rangle \\
\nonumber{}-  12 \langle I (t + \tau) I (t) I (t - \tau) \rangle +
12 \langle I (t + \tau) I^2 (t) \rangle
 \end{gather}
and define the second-order asymmetry function as
\begin{gather}
\gamma_2 (\tau)  = {-} \frac{\langle [\Delta_2 (\tau)]^3
\rangle}{\langle [\Delta_2 (\tau)]^2 \rangle^{3/2}} .  \label{eq8:Shishov_n}
\end{gather}
The definition of the function $\gamma_2 (\tau)$ formally corresponds
to the definition of the asymmetry coefficient for a random
quantity, $\Delta_2 (\tau)$. Since the third moment of $\Delta_2 (\tau)$
is negative in the case of scintillations, we have added
a minus sign in (8), in order for the function  $\gamma_2 (\tau)$ to be
positive for intensity fluctuations due to scintillation.

We also introduce the somewhat different asymmetry function
\begin{gather}\label{eq9:Shishov_n}
\gamma_{2,1} (\tau)  = {-} 2   \frac{\langle [\Delta_2 (\tau) ]^3
\rangle}{\langle[\Delta_2 (\tau) ]^2\rangle
\langle[\Delta_1 (2\tau) ]^2\rangle^{1/2}} \\
\nonumber{}= {-} 2  \frac{\langle[\Delta_2 (\tau) ]^3\rangle}{[4
D_I (\tau)  -
 D_I (2\tau)] [D_I (2\tau) ]^{1/2}} .
\end{gather}
As will be shown below in Appendices A and B, in the case of
intensity fluctuations distributed according to a Rice--Nakagami law or
lognormal law, the function $\gamma_{2,1} (\tau)$ for small values of
$\tau$ is related to the function ${[ D_I (2\tau)]^{1/2}}/{\langle I
\rangle}$ by a formula of the form (2):
\begin{gather}
\gamma_{2,1} (\tau)  = \frac{A [D_I (2\tau)]^{1/2}}{\langle I
\rangle} .
 \label{eq10:Shishov_n} \end{gather}
In general, the factor $A$ is a function of $\tau$, and we
introduce the function $A_{2,1} (\tau)$:
\begin{gather}
A_{2,1} (\tau)  = \langle I \rangle \frac{\gamma_{2,1}
 (\tau)} {[D_I (2\tau)]^{1/2}} . \label{eq11:Shishov_n}
\end{gather}
The form of the functions $\gamma_2 (\tau)$, $\gamma_{2,1} (\tau)$,
and $A_{2,1} (\tau)$ for the two distribution laws (a normal
distribution for wave-field fluctuations and a lognormal law for
intensity fluctuations) is determined in Appendices A and B.

\section{WEAK SCINTILLATIONS}

\begin{figure}[]
\includegraphics[angle=0,scale=0.4]{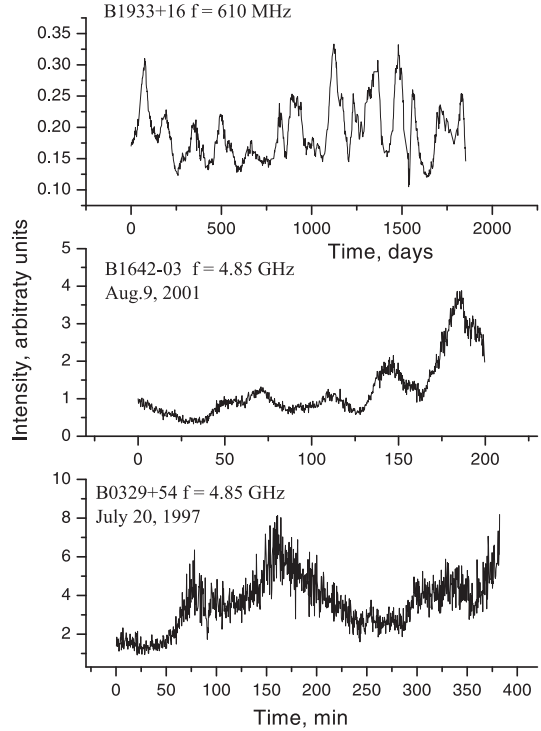}
\setcaptionmargin{5mm}
\onelinecaptionsfalse  
\caption{Time variations of the emission intensity for
PSR~B1933$+$16 at 610~MHz (top), PSR~B1642$-$03 at 4.85 GHz (center),
and PSR~B0329$+$54 at 4.85~GHz (bottom). \hfill}
\end{figure}

We will consider the properties of the asymmetry function in the
case of scintillations of radio sources on electron-density
inhomogeneities of plasma with a power-law turbulence spectrum.
It is known that, in this case, the scintillation regimes are determined
by the structure function of the phase fluctuations on a
scale that is equal to the scale of the first Fresnel zone $D_S
(b_{Fr})$, where $b_{Fr} = \left({r_{eff}}/{k}\right)^{1/2}$,
$r_{eff}$ is the effective distance from the observer or the
source to the layer containing the turbulent medium (in the case of a
statistically uniform medium, this is the distance from the
observer to the source), and $k = {2\pi}/{\lambda}$ is the
wavenumber. If
\begin{gather}
D_S (b_{Fr}) \ll  1, \label{eq12:Shishov_n}
\end{gather}
the scintillations are weak, the characteristic spatial scale giving rise
to the intensity fluctuations is equal to $b_{Fr}$, and the
scintillation index can be found from the formula [9]
\begin{gather}
m^2 \cong K(n) D_S (b_{Fr}), \label{eq13:Shishov_n}
\end{gather}
where $K(n)$ is a numerical factor of the order of unity that
depends on the exponent $n$ of the turbulence spectrum.

Numerical calculations of the one-dimensional distribution function
of the flux fluctuations $F(I)$ were carried out in [10, 11], and the
second and third moments of the intensity fluctuations in
the case of a Kolmogorov spectrum for the inhomogeneities of the
index of refraction for various values of the inner turbulence
scale were determined. It was shown that, in the weak
scintillation regime, as well in the strong scintillation regime with
values of $D_S (b_{Fr})$ that are not too large, the distribution function
$f(I)$ is close to a lognormal function. Using the results of
these studies, we find $A=2.78$ for a
Kolmogorov spectrum without a turnover and $A=2.86$ for a
Kolmogorov spectrum with an inner scale $l$ equal to the scale $b_F$
of the first Fresnel zone (${l}/{b_F} = 1$). Thus, with
increase in the inner scale of the turbulence spectrum, the factor $A$
increases and approaches three, which corresponds to a lognormal
distribution for the flux fluctuations. The calculations show that
relationship (2) is valid up to a value of $m$ that
approaches unity from the side of the weak (unsaturated)
scintillation regime.

\begin{figure}[]
\includegraphics[angle=0,scale=0.4]{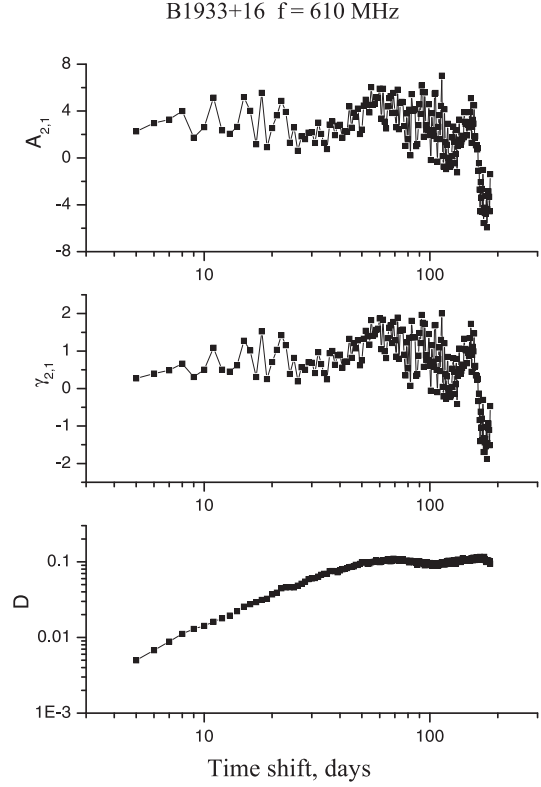}
\setcaptionmargin{5mm}
\onelinecaptionsfalse  
\caption{Structure function, asymmetry function, and coefficient
$A_{2,1}$ versus the time shift for refractive scintillations of
PSR~B1933$+$16. \hfill}
\end{figure}

Note that the one-dimensional distribution function
$f(I)$ and asymmetry coefficient $\gamma$ describe the flux
fluctuations on the main scale of the spatial pattern of
the scintillations and, accordingly, on the main temporal scale
$\tau_0$ of the structure function of the flux fluctuations. When
$\tau \ge \tau_0$, the functions $\gamma_2 (\tau)$, $\gamma_{2,1}
(\tau)$, and $A_{2,1} (\tau)$ will be determined by asymptotic
relationships that correspond to a lognormal distribution law
for the intensity fluctuations (see Appendix B):
\begin{gather}
\gamma_2 (\tau)  \cong   \frac{3}{2}\, \frac{D_I (2\tau)}{[4D_I
(\tau) - D_I (2\tau)]^{1/2} \langle I\rangle} ,
\label{eq14:Shishov_n}
\end{gather}
\begin{gather}
\gamma_{2,1} (\tau)  \cong  3[D_I (2\tau)]^{1/2}/{\langle
I\rangle}, \label{eq15:Shishov_n} \end{gather}
\begin{gather}
A_{2,1} (\tau)  \cong  3 . \label{eq16:Shishov_n}
\end{gather}

Flux fluctuations with small spatial and temporal scales are
determined by the diffraction of the radio waves on inhomogeneities
in the index of refraction with sizes that are much smaller than the
first Fresnel zone. Therefore, the distribution of small-scale
flux fluctuations will more closely fit a normal distribution law,
and the functions $\gamma_2 (\tau)$, $\gamma_{2,1}
(\tau)$, and $A_{2,1} (\tau)$ will better match the asymptotic
relationships (A.8)--(A.10) of Appendix A.

\section{DIFFRACTIVE SCINTILLATIONS}

\begin{figure}[]
\includegraphics[angle=0,scale=0.4]{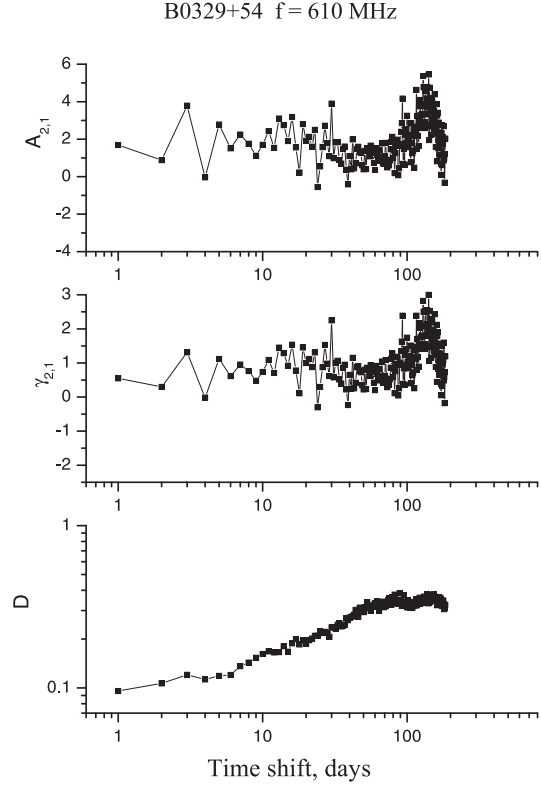}
\setcaptionmargin{5mm}
\onelinecaptionstrue  
\caption{Same as Fig.~2 for PSR~B0329$+$54. \hfill}
\end{figure}

For large values of the parameter $D_S (b_{Fr})$, the intensity
fluctuations consist of two components: small-scale (diffractive)
and large-scale (refractive). With increasing $D_S
(b_{Fr})$, the distribution function of the field of the
diffractive component tends to a normal distribution [9], the
scintillation index tends to unity, and the characteristic
spatial scale of the diffraction pattern corresponds to the field
coherence scale ${1}/{k \theta_{scat}}$, where $\theta_{scat}$ is
the characteristics scattering angle. When $\tau \le \tau_0$,
the functions $\gamma_2 (\tau)$, $\gamma_{2,1} (\tau)$, and $A_{2,1}
(\tau)$ will correspond to the asymptotic expressions (Appendix A)
\begin{gather}
\gamma_2 (\tau) \cong \frac{3}{4}\, \frac{D_I (2\tau)}{[4D_I
(\tau) - D_I (2\tau)]^{1/2} \langle I\rangle} ,
\label{eq17:Shishov_n}
\end{gather}
\begin{gather}
\gamma_{2,1} (\tau) \cong \frac{3}{2} \frac{[D_I
(2\tau)]^{1/2}}{\langle I\rangle} ,  \label{eq18:Shishov_n}
\end{gather}
\begin{gather}
A_{2,1} (\tau)  = \frac{\gamma_{2,1} (\tau)}{[D_I (2\tau)]^{1/2}}
\cong \frac{3}{2} , \label{eq19:Shishov_n}
\end{gather}
and, when $\tau \gg \tau_0$, will be determined as $\gamma_2
(\tau) \cong  2 / \sqrt{6}$, $\gamma_{2,1} (\tau) \cong 2
\sqrt{2}$, $A_{2,1} (\tau) \cong  2$.

\section{REFRACTIVE SCINTILLATIONS}

\begin{figure}[]
\includegraphics[angle=0,scale=0.5]{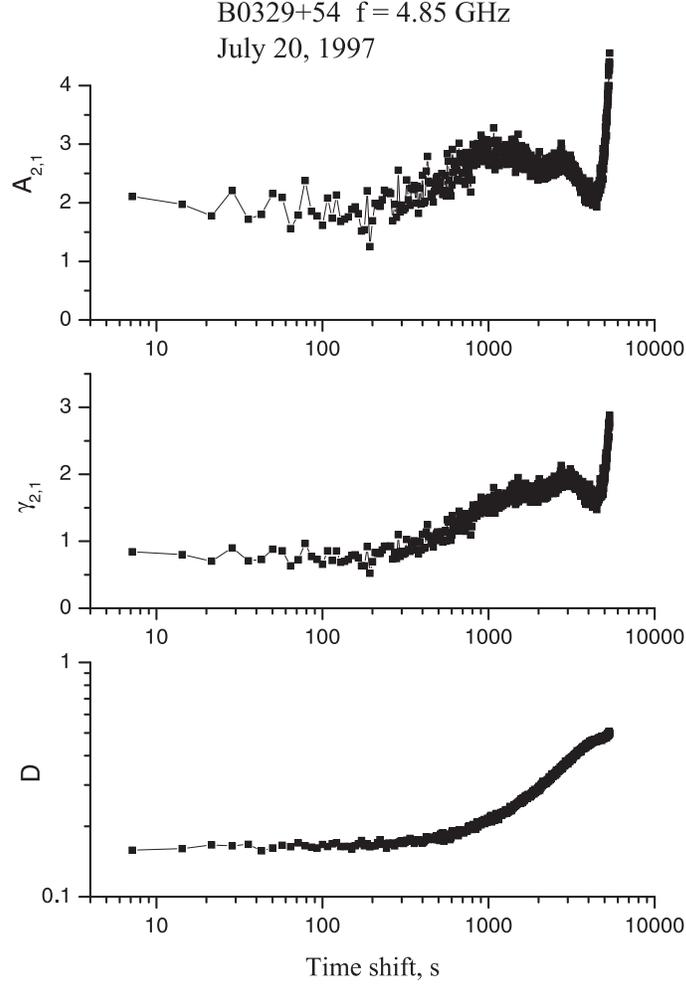}
\setcaptionmargin{5mm}
\onelinecaptionstrue  
\caption{Same as Fig.~3 but for weak scintillations. \hfill}
\end{figure}

The refractive component is determined by weak focusing of
the radiation on inhomogeneities with characteristic scales of the
order of the scattering disk: $r_{eff} \theta_{scat}$ [9].
Refractive scintillations can be treated as weak scintillations
on large-scale inhomogeneities, with sizes much larger than the scale of the
first Fresnel zone. Therefore, we expect a lognormal
distribution law for the flux fluctuations. When $\tau \ge \tau_0$,
the functions $\gamma_2 (\tau)$, $\gamma_{2,1} (\tau)$, and $A_{2,1}
(\tau)$ will be determined by the asymptotic relationships (14)--(16).

\section{COMPARISON WITH OBSERVATIONS}

We used data on diffractive, refractive, and weak scintillations of
several pulsars to analyze the applicability of the above relationships
for the asymmetry function to actual observations of interstellar
scintillations of pulsars.

\subsection*{Refractive and Weak Scintillations}

\begin{figure}[]
\includegraphics[angle=0,scale=0.4]{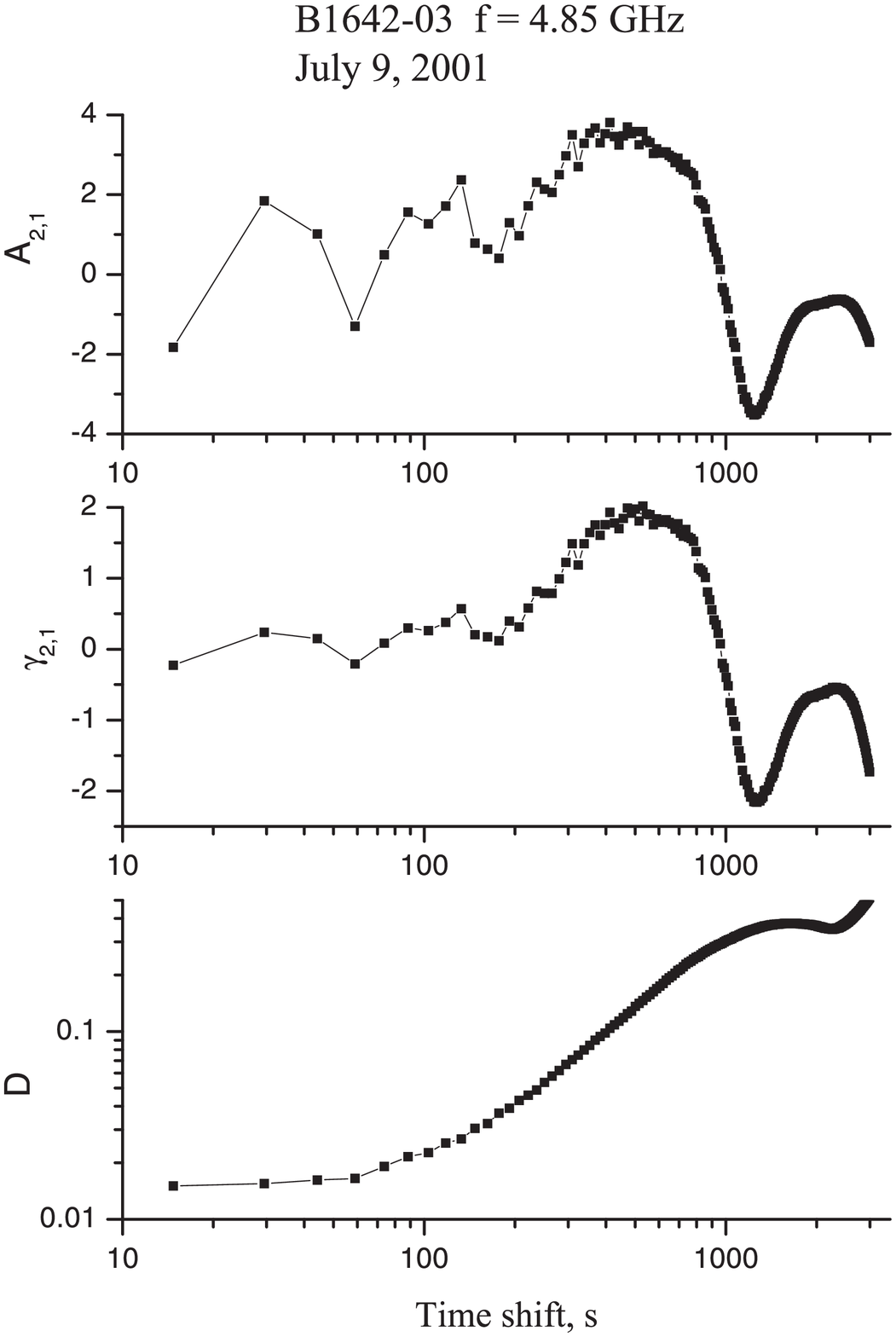}
\setcaptionmargin{5mm}
\onelinecaptionstrue  
\caption{Same as Fig.~4 but for PSR~B1642$-$03. \hfill}
\end{figure}

We used the five-year monitoring data for PSR~В1933$+$16 and В0329${+}$54
(1990--1995) at 610~MHz [12] (NRAO, Green Bank) to analyze refractive
scintillations. The observations were carried out in a 16-MHz band;
diffractive scintillations were considerably smoothed by
averaging the signal over about an hour of observations. Figure~1
(top) shows the time variations of the emission intensity for
PSR~В1933$+$16. The time scale of these variations is 33~days
[12]. We used relationships (5), (9), and (11) to calculate
the structure function of the flux fluctuations $D(\tau)$, asymmetry
function $\gamma_{2,1} (\tau)$, and coefficient $A_{2,1} (\tau)$
for all the types of scintillations considered below. Before
calculating these functions, we have normalized the time series to
their average values.

Figures 2 and 3 show $D(\tau)$, $\gamma_{2,1} (\tau)$, and
$A_{2,1} (\tau)$ as functions of the time shift in days for
PSR~В1933$+$16 and В0329${+}$54, respectively. The structure
function reaches a constant level equal to $2m^2$, where
$m$ is the modulation index of the intensity variations. The time
shift at which $D(\tau)$ decreases by a factor or two gives the
time scale of these variations. In $\gamma_{2,1} (\tau)$
and $Ac (\tau)$, this time scale corresponds to a shift that is a factor
of two smaller [see formula (10)]. We can see from Figs. 2 and
3 that $A_{2,1} (\tau)$ is approximately constant for shifts of
less than 40~days, and this average level is $A_{2,1}\approx
3$ for PSR~В1933$+$16, based on formula (5) for refractive
scintillations, and $A_{2,1}\approx 2$ for PSR~B0329$+$54. The
difference of $A_{2,1}$ from three for PSR~B0329$+$54 may be due to
the fact that the index of the electron-density fluctuation
spectrum toward this pulsar differs from the Kolmogorov index~[13]:
it is $n=3.5$. It seems reasonable to suppose that the decrease of
the spectral index should result in a decrease in $A_{2,1}$ at
small time shifts.

\begin{figure}[]
\includegraphics[angle=0,scale=0.4]{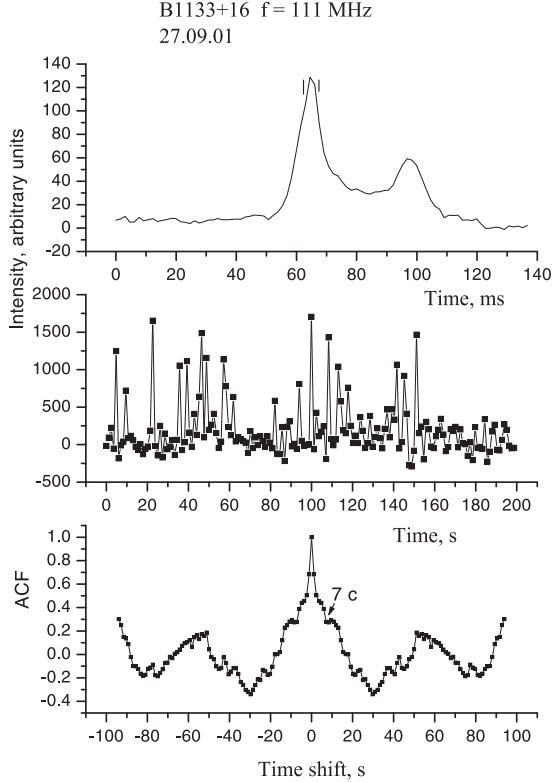}
\setcaptionmargin{5mm}
\onelinecaptionsfalse  
\caption{Top: average profile of PSR~B1133$+$16 at 111~MHz.
Center: pulse-to-pulse intensity variations in a single 1.25-kHz
channel (for each pulse the intensity was obtained by integrating
within the longitude interval shown with dashes on the average
profile). Bottom: autocorrelation function of these intensity
variations. \hfill}
\end{figure}

We used observations of the pulsars PSR~В0329$+$54 and
В1642$-$03 carried out at 4.85~GHz on the 100-m Effelsberg telescope for
our analysis of weak scintillations. The input data and observing
parameters are given in [13, 14]. Before recording the signal,
we averaged the pulsed emission over $10\,P_1$ for PSR~В0329${+}$54
and $39\,P_1$ for PSR~В1642$-$03 (where $P_1$ is the
pulsar period). The time resolutions were 7.15 and 15~s,
respectively. Figure~1 shows the time series for these pulsars
(central and bottom graphs). We used the fourth channel (the
observations were carried out in four frequency channels of 60~MHz
each), which was the most free of interference, for PSR~В0329${+}$54.
The modulation indices for the intensity variations for
PSR~В0329${+}$54 and В1642$-$03 were 0.45 and 0.64, respectively.

Figures 4 and 5 show $D(\tau)$, $\gamma_{2,1} (\tau)$, and
$A_{2,1} (\tau)$ as functions of the time shift in seconds. The
maximum time shift for which these functions were calculated was
one-fourth of the total time interval of the observations. The
scintillation time scales for PSR~В0329${+}$54 and В1642$-$03 are 2200 and
1170~s, respectively (at the $1/e$ level of their
autocorrelation functions). On time scales of 1000--2000~s for
PSR~В0329${+}$54 and 400--700~s for PSR~В1642$-$03, the values of
$A_{2,1} (\tau)$ are 2.8 and 3.3, respectively. These agree
well with the value $A_{2,1} = 3$ predicted by the theory for
weak pulsations [formula (16)]. The constant level at the
shortest time shifts in the structure function $D(\tau)$ is
due to uncorrelated noise.

We did not correct the calculated functions for noise, because, as
was shown by the analysis, their third moments are nonzero, and
this correction would introduce strong distortions to $\gamma_{2,1}
(\tau)$ and $A_{2,1} (\tau)$. The constant level $A_{2,1}\approx
2$ at time shifts $\tau\le 200$~s for PSR~В0329$+$54 is probably
due to diffractive scintillations, whose time scale at 610~MHz is
$t_d \approx 90$~s [13]. For PSR~В1642$-$03, $t_d \approx 130$~s
at the same frequency [14]. For this pulsar there are strong
fluctuations in $A_{2,1} (\tau)$ at shifts shorter than 200~s,
due to the effect of noise; however, the mean level
of about 1.5 is probably associated with diffractive
scintillations on small-scale inhomogeneities.

\subsection*{Diffractive Scintillations}

We used observations of PSR~B1133$+$16 carried out
on September 27 and 29, 2001 on the Large Phased Array of the Lebedev
Physical Institute in Pushchino at 111~MHz to analyze the intensity
variations of this pulsar using the asymmetry function.
Individual pulses of the pulsar were recorded in 96 channels of
the receiver (the bandwidth of each channel was 1.25~kHz) with a
time resolution of 1.56~ms during 200~s, and were stored on a
computer disk. We performed the structural analysis after removing
the dispersion shift in all frequency channels and integrating the
spectrum in the selected interval of pulse longitudes. Figure~6 (top)
shows the mean pulse of  PSR~B1133$+$16,
obtained by accumulating pulses during 200~s in a
120-kHz band. The same figure (center) presents the pulse-to-pulse
intensity variations in one of the channels (1.25~kHz). The
intensity was obtained for each pulse in all the channels by
integrating the signal in the longitude interval of the first
component of the mean profile, which is denoted in Fig.~6 with
dashes. The same figure (bottom) shows the autocorrelation
function (ACF) calculated for the intensity fluctuations
$I(t)$ in this channel. The ACF has three time scales: the first
is due to pulse-to-pulse variations within one pulsar period
($P_1 = 1.188$~s), the secondm which is about 7~s, is
due to diffractive scintillations, and the third represents slower
variations on a time scale of $\sim$20~s. The fact that the 7-s time scale is
due to diffractive scintillations follows from the fact that
the decorrelation of the spectra (96 channels 1.25~kHz each) occurs
precisely on this time scale (with pulses separated by $6\,P_1$).
To reduce the effect of intrinsic rapid variations of the pulsar,
we smoothed the input $I(t)$ data in all the channels on a
three-point interval ($3\,P_1$). When calculating $D(\tau)$,
$\gamma_{2,1} (\tau)$, and $A_{2,1} (\tau)$, we averaged the
second and third moments of $I(t)$ over all frequency channels.
Figure~7 shows the obtained functions averaged over two
observational days. We can see that the different
time scales are well separated in the functions $\gamma_{2,1} (\tau)$
and $A_{2,1} (\tau)$. The value of $A_{2,1} (\tau)$ on scales
$\leq 7$~s is 1.4, in good agreement with the theoretical value of
1.5, corresponding to diffractive scintillations.

\begin{figure}[]
\includegraphics[scale=0.4]{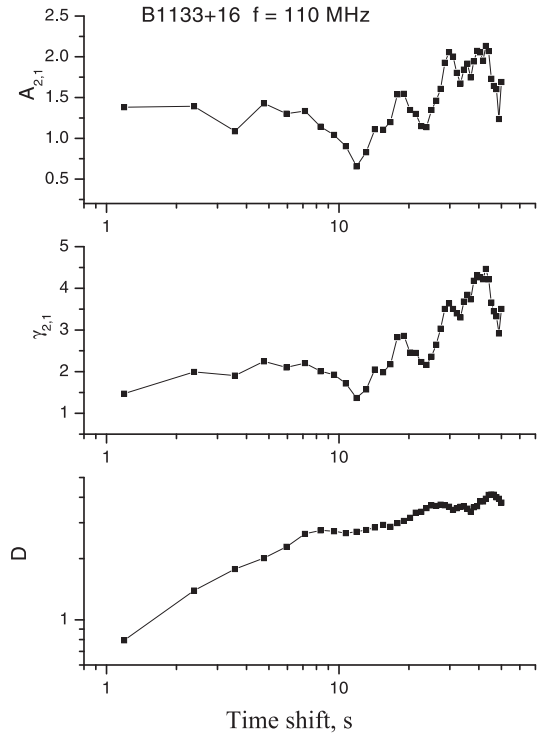}
\setcaptionmargin{5mm}
\onelinecaptionsfalse  
\caption{Structure function, asymmetry function, and coefficient
$A_{2,1}$ versus time shift for PSR~B1133$+$16 at 111~MHz
averaged over two observational days (diffractive
scintillations). \hfill}
\end{figure}

\section{CONCLUSION}

Our analysis of observed pulsar intensity variations
using the asymmetry function $\gamma_{2,1}(\tau)$ and function
$A_{2,1}(\tau)$ has shown that these variations match the behavior
expected for a normal distribution for the field fluctuations
(diffractive scintillations) or a lognormal distribution for the intensity
fluctuations (refractive scintillations, weak scintillations), in
accordance with the derived theoretical relationships. For all
the scintillation regimes, there is a functional relationship between
the asymmetry function and the structure function of the intensity
fluctuations. This relationship has the simplest form for the asymmetry
function $\gamma_{2,1}(\tau)$ at small values of $\tau$:
$\gamma_{2,1}(\tau) \propto \sqrt{D_{2,1}(2\tau)}$. Accordingly,
the function $A_{2,1} (\tau) = \langle I \rangle
\dfrac{\gamma_{2,1}(\tau)}{\sqrt{D_I (2\tau)}}$ is a constant.
This indicates that analysis of the function
$A_{2,1}(\tau)$ could provide a good test for isolating scintillations
against the background of variations that have other origins,
and that the proposed method can be efficiently used to
study and separate out intensity variations that have different
physical natures. In particular, it can be applied to the analysis
of flux variations of extragalactic sources, in order to distinguish
variations due to intrinsic variability of the source and effects
associated with the propagation of the radiation in the interstellar
plasma.

We note also that there are prospects for further studies
in this direction. It is necessary to study the dependence of the
form of the asymmetry function on the spectrum of the inhomogeneities
in the index of refraction, in particular, for a power-law spectrum, and
the dependence of the asymmetry function on the spectral index.
Studies of the dependence of the form of the asymmetry function on
the source structure are also required. If such theoretical
calculations, confirmed by experimental data, are available,
measurements of the asymmetry function can yield additional
information about both the medium and source.

\section*{ACKNOWLEDGMENTS}

This work was supported by the Russian Foundation for Basic
Research (project codes 03-02-16522, 03-02-16509), INTAS
(grant 00-849), the NSF (grant AST 0098685), the Federal Science
and Technology Program in Astronomy, and the Program of the Division
of Physical Sciences of the Russian Academy of Sciences on ``Fundamental
Studies of Extended Objects in the Universe.''

\begin{flushright}
\emph{Appendix A}
\end{flushright}
\section*{THEORETICAL RELATIONSHIP BETWEEN THE ASYMMETRY
FUNCTIONS $\gamma_2 (\tau)$, $\gamma_{2,1} (\tau)$ AND THE STRUCTURE
FUNCTION $D_I (\tau)$ FOR A NORMAL DISTRIBUTION FOR THE FIELD
FLUCTUATIONS}

Let us assume that the wave-field fluctuations have a normal
distribution and that the average value of the field is zero:
\begin{gather*}\tag{\mbox{A}.1}
\langle E (t) \rangle =  0.
\end{gather*}
This equality implies that second-order moments of the form
$\langle E^2(t_1)\rangle$ are also equal to zero [9]:
\begin{gather*}\tag{\mbox{A}.2}
\langle E^2 (t_1) \rangle =  \langle [ E^* (t_1) ]^2 \rangle  =  0.
\end{gather*}

Formulas (A.1) and (A.2) correspond to diffractive scintillations
in the saturated scintillation regime [9]. In this case, the third
moment of the intensity or the sixth moment of the field can be represented
as a sum of products of second-order field-coherence functions
[15]:
\begin{gather*}\tag{\mbox{A}.3}
\langle I (t_1) I (t_2) I(t_3) \rangle \\
\nonumber{}=  \langle E (t_1) E^* (t_1)
                            E (t_2) E^* (t^2) E (t_3) E^* (t_3) \rangle \\
                            \nonumber{}=
\langle E (t_1) E^* (t_1) \rangle
\langle E (t_2) E^* (t_2) \rangle
\langle E (t_3) E^* (t_3)\rangle \\
\nonumber{}+ \langle E (t_1) E^* (t_2) \rangle \langle E (t_2) E^*
(t_1) \rangle
\langle E (t_3) E^* (t_3) \rangle \\
\nonumber{}+ \langle E (t_1) E^* (t_3) \rangle \langle E (t_2) E^*
(t_2) \rangle
\langle E (t_3) E^* (t_1) \rangle \\
\nonumber{}+ \langle E (t_1) E^* (t_1) \rangle \langle E (t_2) E^*
(t_3) \rangle
\langle E (t_3) E^* (t_2) \rangle \\
\nonumber{}+ \langle E (t_1) E^* (t_2) \rangle \langle E (t_2) E^*
(t_3) \rangle
\langle E (t_3) E^* (t_2) \rangle \\
\nonumber{}+ \langle E (t_1) E^* (t_3) \rangle \langle E (t_2) E^*
(t_1) \rangle
\langle E (t_3) E^* (t_2) \rangle \\
\nonumber{}= \langle I \rangle^3 + \langle I \rangle
[B_I (t_1 -t_2) + B_I (t_1 - t_3) \\
\nonumber{}+ B_I (t_2 - t_3 )]  +
B_E (t_1 - t_2) B_E (t_2 - t_3) \\
\nonumber{}\times B_E (t_3 - t_2)  +  B_E (t_1 - t_3) B_E (t_2 -
t_1) \\
\nonumber{}\times B_E (t_3 - t_2) .
\end{gather*}
Here,
\begin{gather*} \tag{\mbox{A}.4}
B_E (t_i - t_k) = \langle E (t_i) E^* (t_k) \rangle
\end{gather*}
is the field-coherence function and $B_I (t_i - t_k) =  \langle
[ I (t_i) - \langle I (t_i) \rangle ] [I (t_k) - \langle
I(t_k)\rangle ] \rangle = B_E (t_i - t_k)\times B^*_E (t_i -
t_k)$ is the correlation function of the intensity fluctuations.
Using formula (A.3) in the calculation in (7), we
obtain a relationship for the third moment of the second
difference:
\begin{gather*}\tag{\mbox{A}.5}
\langle [\Delta_2 (\tau)]^3 \rangle
 = {-} 12 \langle I \rangle^3 + 12 \langle I \rangle
B_I (2\tau) \\
\nonumber{}+ 24 \langle I \rangle B_I (\tau) - 24 B_E (2\tau) B_I
(\tau)  ={-} 6 \langle I \rangle D_I (2\tau) \\
\nonumber{}+ 24 \langle I \rangle \bigg\{\langle I \rangle -
\bigg[\langle I \rangle^2 - \frac{1}{2}
D_I (2\tau)\bigg]^{1/2}\bigg\} \\
\nonumber{}-  12 D_I (2\tau)\bigg\{ \langle I \rangle - \bigg[
\langle I \rangle^2 - \frac{1}{2} D_I (2\tau)\bigg]^{1/2} \bigg\}
.
\end{gather*}

Let $\tau_0$ be the characteristic scale of the structure
function of the intensity fluctuations $D_I (\tau)$. For small
values, $\tau \le \tau_0$, we obtain the asymptotic formula
\begin{gather*}\tag{\mbox{A}.6}
\langle [\Delta_2 (\tau)]^3 \rangle \cong \frac{3}{4} D_I (2\tau)
\frac{4 D_I (\tau) - D_I (2\tau)}{\langle I \rangle} .
\end{gather*}
For large values, $\tau \gg \tau_0$, using equality  $D_I (\tau)
= 2 \langle(I- \langle I \rangle)^2 \rangle = 2 \langle I
\rangle^2$ yields
\begin{gather*}\tag{\mbox{A}.7}
\langle [\Delta_2 (\tau)]^3 \rangle \cong - 12 \langle I \rangle^3
. \end{gather*}

Using these asymptotic relationships for $\langle [\Delta_2
(\tau)]^3 \rangle$ and formulas (6) and (8), we derive for the function
$\gamma_2 (\tau)$ the asymptotic expressions
\begin{gather*}\tag{\mbox{A}.8}
\gamma_2 (\tau)\cong \begin{cases}
   \dfrac{3}{4} \, \dfrac{D_I
(2\tau)}{[4D_I (\tau) - D_I (2\tau)]^{1/2} \langle I \rangle}, \quad \tau \le \tau_0 , \\
               \dfrac{2}{\sqrt{6}} , \quad \tau \gg \tau_0 .
\end{cases}  \end{gather*}

Similarly, we have
\begin{gather*}\tag{\mbox{A}.9}
\gamma_{2,1} (\tau)
\cong
\begin{cases}
 \dfrac{3}{2}[D_I
             (2\tau)]^{1/2}/ \langle I \rangle, \quad \tau \le \tau_0  , \\
2 \sqrt{2}, \quad \tau \gg \tau_0 .
\end{cases}
\end{gather*}
We see that, for small values of $\tau$, $\gamma_{2,1}
(\tau)$ is related to ${[D_I (2\tau)]^{1/2}}/{\langle I \rangle}$
by the same formula as that relating the asymmetry coefficient to the
scintillation index.

We obtain for $A_{2,1} (\tau)$
\begin{gather*}\tag{\mbox{A}.10}
A_{2,1} (\tau)   =   \langle I \rangle \frac{\gamma_{2,1}
(\tau)}{[D_I (2\tau)]^{1/2}} \cong \frac{3}{2} , \quad \tau \le
\tau_0  , \\ A_{2,1} (\tau)  \cong  2 , \quad \tau \gg \tau_0 .
\end{gather*}

\begin{flushright}
\emph{Appendix B}
\end{flushright}

\section*{THEORETICAL RELATIONSHIP BETWEEN THE ASYMMETRY FUNCTIONS
$\gamma_2 (\tau)$, $\gamma_{2,1} (\tau)$ AND THE STRUCTURE FUNCTION
$D_I (\tau)$ FOR A LOGNORMAL DISTRIBUTION FOR THE INTENSITY FLUCTUATIONS}

Let us represent the intensity $I(t)$ as
\begin{gather*}\tag{\mbox{B}.1}
I (t)  =  I_0 \exp [\chi (t)]
\end{gather*}
and assume that the fluctuation distribution law $\chi(t)$ is normal.
For a normally distributed quantity, the following equality is
valid [8]:
\begin{gather*}\tag{\mbox{B}.2}
I_0 \langle \exp[\chi (t)] \rangle = I_0 \exp\left[ \langle \chi
(t) \rangle + \frac{1}{2} \langle( \delta \chi (t))^2
\rangle\right].
\end{gather*}
Taking into account the fact that the mean intensity does not change
during the propagation of the wave in a turbulent medium, i.e.,
$\langle I \rangle = I_0$, we obtain
\begin{gather*}\tag{\mbox{B}.3}
\langle \chi (t) \rangle  = {-} \frac{1}{2} \langle ( \delta \chi
(t))^2 \rangle .
\end{gather*}

We find with (B.2) and (B.3)
\begin{gather*}\tag{\mbox{B}.4}
\langle I^2 \rangle = \langle I_0^2 \exp[2\chi (t)]\rangle \\
\nonumber{}= I_0^2 \exp\left[ 2 \langle \chi (t) \rangle  + 2
\langle (\delta \chi (t))^2 \rangle \right] \\
\nonumber{}= I_0^2 \exp\left[ \langle (\delta \chi (t))^2 \rangle
\right],
\end{gather*}
\begin{gather*}\tag{\mbox{B}.5}
\langle I (t) I(t+\tau) \rangle  = \exp\left[ \langle \delta
\chi(t) \delta \chi (t+\tau) \rangle \right] \\
\nonumber{}= \langle I^2 \rangle \exp\left[-\frac{1}{2} D_{\chi}
(\tau)\right],
\end{gather*}
where
\begin{gather*} \tag{\mbox{B}.6}
D_{\chi} (\tau)  = \langle
\left[\chi (t+\tau) - \chi (t)\right]^2 \rangle \\
\nonumber{}= 2 \langle \left[\delta \chi (t)\right]^2\rangle - 2
\langle \delta \chi (t) \delta \chi (t+\tau)\rangle
\end{gather*}
is the structure function of the fluctuations, $\chi (t)$.

Formulas (B.4) and (B.5) enable us to express the quantities
$\exp\left[ \langle (\delta \chi (t))^2\rangle\right]$ and
$\exp\left[-\frac{1}{2} D_{\chi} (\tau)\right]$ in terms of the
scintillation index and the structure function of the intensity
fluctuations:
\begin{gather*}\tag{\mbox{B}.7}
\exp[ \langle (\delta \chi (t))^2\rangle]  = 1 +  m^2 ,
\end{gather*}
\begin{gather*}\tag{\mbox{B}.8}
\exp\left[-\frac{1}{2} D_{\chi} (\tau)\right]  = 1 - \frac{D_I
(\tau)}{2 I_0^2 (1+m^2)} . \end{gather*}
Then, using the formula
\begin{gather*}\tag{\mbox{B}.9}
\langle I (t_1) I(t_2) I (t_3)\rangle \\
\nonumber{}=  I_0^3 \exp\bigg[3\langle (\delta \chi)^2\rangle -
\frac{1}{2} D_{\chi} (t_1-t_2) \\
\nonumber{}-\frac{1}{2} D_{\chi} (t_1-t_3) - \frac{1}{2} D_{\chi}
(t_2-t_3)\bigg]   = I_0^3 (1+m^2)^3 \\
\nonumber{}\times\exp\bigg[- \frac{1}{2} [D_{\chi} (t_1-t_2) +
D_{\chi} (t_1-t_3) \\
\nonumber{}+ D_{\chi} (t_2-t_3)]\bigg],
\end{gather*}
we obtain for the third moment of the
second difference $\Delta_2 (\tau)$
\begin{gather*}\tag{\mbox{B}.10}
\langle [\Delta_2 (\tau)]^3 \rangle  = {-} 6 I_0^3 \exp\left[ 3
\langle (\delta \chi)^2 \rangle\right] \\
\nonumber{}\times\bigg\{1- \exp[- D_{\chi} (2\tau)] -
2\exp[-D_{\chi} (\tau)] \\
\nonumber{}+2\exp\left[- D_{\chi} (\tau) - \frac{1}{2}
D_{\chi}(2\tau)\right]\bigg\} .
\end{gather*}

Using (B.4) and (B.5), we find
\begin{gather*}\tag{\mbox{B}.11}
\langle [\Delta_2 (\tau)]^3 \rangle  =  {-} 6 I_0^3
(1+m^2)^3 \\
\nonumber{}\times \bigg\{1 - \left[1- \frac{D_{\chi} (2\tau)}{2
I_0^2 (1+m^2)}\right]^2 + 2\,\frac{D_{\chi} (2\tau)}{I_0^2
(1+m^2)} \\
\nonumber{}\times\left[1- \frac{D_{\chi} (\tau)}{2 I_0^2
(1+m^2)}\right]^2\bigg\}.
\end{gather*}
For small $m^2 \ll 1$, we obtain the asymptotic relationship
\begin{gather*}\tag{\mbox{B}.12}
\langle [\Delta_2 (\tau)]^3 \rangle \cong \frac{3}{2} D_I (2\tau)
\frac{4 D_I (\tau) - D_I (2\tau)}{I_0} .
\end{gather*}
Substituting this asymptotic relationship into (8)--(10) yields
for $m^2 \ll 1$
\begin{gather*}\tag{\mbox{B}.13}
\gamma_2 (\tau)  \cong   \frac{3}{2}\, \frac{D_I (2\tau)}{[4D_I
(\tau) - D_I (2\tau)]^{1/2}I_0} ,
\end{gather*}
\begin{gather*}
\tag{\mbox{B}.14} \gamma_{2,1} (\tau)  \cong  3 \left[D_I
(2\tau)\right]^{1/2}/I_0 ,
\end{gather*}
\begin{gather*}
\tag{\mbox{B}.15} A_{2,1} (\tau)  \cong  3 .
\end{gather*}

Translated by G. Rudnitski{\u\i}

\end{document}